\documentclass[12pt]{article}

\begin{document}

\title{An efficient prescription to find the eigenfunctions of point
interactions Hamiltonians}
\author{F. A. B. Coutinho$^{1}$ and M. Amaku$^{2}$ \\
$^{1}$Faculdade de Medicina, Universidade de S\~{a}o Paulo \\
e LIM 01-HCFMUSP, 05405-000, S\~{a}o Paulo, SP, Brasil \\
$^{2}$Faculdade de Medicina Veterin\'{a}ria e Zootecnia, \\
Universidade de S\~{a}o Paulo, 05508-970, S\~{a}o Paulo, SP, Brasil }
\date{ }
\maketitle



\begin{abstract}
A prescription invented a long time ago by Case and Danilov is used to get
the wave function of point interactions in two and three dimensions.
\end{abstract}


\section{Introduction}

\bigskip Consider the free particle $s$-wave time-independent Schr\"{o}%
dinger equation in $D=2$ and $D=3$ dimensions. The radial part of the Schr%
\"{o}dinger equation is given by%
\begin{equation}
-\frac{\hbar ^{2}}{2m}\frac{1}{r^{D-1}}\frac{d}{dr}r^{D-1}\frac{d\psi _{k}(r)%
}{dr}=E\psi _{k}(r)  \label{eq1a}
\end{equation}%
The most general solutions are given by%
\begin{equation}
\psi _{k}(r)=\cos (\eta (k))J_{0}(kr)-\sin (\eta (k))N_{0}(kr)  \label{eq:1}
\end{equation}%
for $D=2$ dimensions and 
\begin{equation}
\psi _{k}(r)=\cos (\eta (k))\frac{\sin (kr)}{r}-\sin (\eta (k))\frac{\cos
(kr)}{r}  \label{eq:2}
\end{equation}%
for $D=3$ dimensions. In equations (\ref{eq:1}) and (\ref{eq:2}), $k=\sqrt{%
\frac{2mE}{\hbar ^{2}}}$, and $J_{0}(kr)$ and $N_{0}(kr)$ in equation (\ref%
{eq:1}) are the Bessel and Neumann functions~\cite{Abramowitz65},
respectively.

One usually disregards the irregular solution on the grounds that it goes to
infinity at the origin. However, although the wave function diverges, the
probability of finding the particle in a small region around the origin is
finite and so, in fact, there is no reason to reject the irregular solution.

The question to be answered is if there exists self-adjoint operators
(Hamiltonians), such that (\ref{eq:1}) and/or (\ref{eq:2}) are
eigenfunctions of such operators. It can be proved~\cite%
{Albeverio88,coutinho04} that for both cases (eigenfunctions (\ref{eq:1})
and (\ref{eq:2})) there exists a family of operators depending on one
parameter that have functions of this form as eigenfunctions. The
eigenfunctions of such operators are given by (\ref{eq:1}) and (\ref{eq:2})
with an appropriate choice of the functions $\eta (k)$ which as we shall see
depend on one parameter that can be taken as, or is related to, the strength
of the interaction. Since the eigenfunctions are indistinguishable from the
free particle eigenfunctions for $r>0$ one usually says that these
Hamiltonians correspond to the free Hamiltonian plus a point interaction at
the origin. Other common name for these interactions are contact
interactions, zero-range interactions or Fermi pseudo potentials.

The subject can be approached from a number of different points of view.
First one can use the theory of self-adjoint extensions~\cite{Albeverio88,
coutinho04, Reed75}. This approach demands a certain mathematical maturity
and although is by far the most complete approach it can hardly be
considered pedagogical for undergraduate students.

Another approach is to use a finite range potential and take the zero range
limit of the potential allowing its strength to diverge suitably. This
regularization procedure has been used for instance in references~\cite%
{Perez91,Coutinho1994,Gosdzinsky91,Mead91}. This process is simple but can
be very laborious.

A third approach is to add Dirac delta function like distributions to the
free Hamiltonian. This approach is considered, for example, in the papers
Kurasov~\cite{kurasov1996,kurasov1997} and again is a little advanced to
undergraduate students.

Some time ago a prescription to handle point interactions was invented by
Danilov~\cite{Danilov61} in the context of many body problems and in a
slightly different context by Case~\cite{case1950}. This prescription was
rediscovered~\cite{Audretsch95} in the context of the scattering of
particles by an Aharonov-Bohm potential.

The purpose of this note is to present Case-Danilov's prescription in the
context of a free particle interacting only with a \textquotedblleft point
interaction\textquotedblright\ in $D=2$ and $D=3$ dimension where it can be
learned easily.

\bigskip Interest in quantum mechanics problems involving point interactions
is continuing since it was introduced by Fermi~\cite{fermi}. A few recent
examples of the its use can be found in the following references~\cite%
{DellAntonio94,Adhikari95,Frederico2006,Girardeau2006,werner2008,Correggi2008}
and in the references therein. The fact that point interactions is so
popular is due to the fact that these problems are frequently solvable~\cite%
{Albeverio88}.

\section{The Case-Danilov prescription}

Case-Danilov's prescription teaches us how to find $\eta (k)$. It consists
in imposing that two eigenfunctions of different energies be orthogonal,
that is 
\begin{equation}
\int_{0}^{\infty }\psi _{k}^{\ast }(r)\psi _{\ell }(r)r^{D-1}dr\quad \propto
\quad \delta (k-\ell )  \label{eq:3}
\end{equation}

For a general $\sin (\eta (k))$ and $\cos (\eta (k))$, the right hand
integral contains terms which are not proportional to $\delta (k-\ell )$. By
choosing $\eta (k)$ in such way that those terms cancel, we get the wave
functions of the self-adjoint family of operators. One should note that
certain integrals that appear below are improper in the ordinary sense and
should be evaluated according, for example, the prescriptions given by
Brownstein~\cite{Brownstein1975}.

This prescription was applied by Case~\cite{case1950} to find the
eigenfunctions of a particle moving under the influence of a potential of
the form $\frac{1}{r^{2}}.$The Case-Danilov prescription can be applied to
find point interactions every time one finds that the time independent Schr%
\"{o}dinger equation has two linear independent solutions that are square
integrable. We mention a few examples at the end of this note.

\section{Calculations}

Consider first the two dimensional case. Using the indefinite integral~\cite%
{Abramowitz65} 
\begin{eqnarray}
\int^{Z} \left[ (k^{2} - \ell^{2})t - \frac{(\mu^{2}-\nu^{2})}{t} \right]
J_{\mu}(kt) N_{\nu}(\ell t) \, dt =  \nonumber \\
= Z [k J_{\mu +1}(kZ) N_{\nu}(\ell Z) - \ell J_{\mu} (kZ) N_{\nu+1} (\ell
Z)] - (\mu - \nu) J_{\mu} (kZ) N_{\nu}(kZ)
\end{eqnarray}
and using the asymptotic forms when $Z \rightarrow 0$ or $Z \rightarrow \infty$
of the Bessel function (\cite{Abramowitz65}, p. 360 for small $Z$ and p. 364 for large $Z$), 
we get (see reference~\cite{Audretsch95} for more details and other cases)
\begin{eqnarray}
\int_{0}^{\infty} \psi_{k}^{*} (r) \psi_{\ell}(r) r dr = \frac{2}{\pi} \tan
(\eta (k)) - \frac{2}{\pi} \tan (\eta (\ell)) +  \nonumber \\
+ \frac{4}{\pi^{2}} \tan ((\eta (k)) \tan (\eta (\ell)) \ln \left( \frac{k}{%
\ell} \right) + \frac{1}{\sqrt{k \ell}} \delta (\ell - k )
\end{eqnarray}
where the terms not proportional to $\delta (k - \ell)$ come from the lower
limit of integration. These terms have to cancel out and this occurs if we
impose 
\begin{equation}
\tan (\eta (k)) = \frac{\pi}{2} \frac{-1}{\ln (k_{b} / k)}  \label{eq:6}
\end{equation}

The physical meaning of the parameter $k_{b}$, which we note is positive,
will be explained below.

Consider now the three dimensional case. Again integrating we find that 
\begin{equation}
\int_{0}^{\infty} \psi_{k}^{*} (r) \psi_{\ell}(r) r^{2} dr = \frac{1}{%
\ell^{2} - k^{2}} [-\tan (\eta (k)) \ell + \tan (\eta (\ell)) k] + \frac{\pi%
}{2} \delta (k - \ell)
\end{equation}
where once again the terms not proportional to $\delta (k- \ell)$ come from
the lower limit of integration. In order to cancel these terms, we impose 
\begin{equation}
\tan (\eta (k)) = - \frac{k}{k_{b}^{\prime}}  \label{eq:8}
\end{equation}
where $k_{b}^{\prime}$ can be positive or negative. This equation can also be derived
from equation (A3) on p. 175 of reference~\cite{Brownstein1975}.

The physical meanings of the parameters $k_{b}$ in equation (\ref{eq:6}) and 
$k_{b}^{\prime }$ in equation (\ref{eq:8}) are as follows. Comparing
equation (\ref{eq:6}) with equation (18) of reference~\cite{Perez91}, we see
that $k_{b}$ is linked with the arbitrary energy of a unique bound state
through $k_{b}=\sqrt{\frac{2m}{\hbar ^{2}}E_{b}}$. The interpretation of $%
k_{b}^{\prime }$ is obtained by observing that $k\cot (\eta
(k))=-k_{b}^{\prime }$ so that $k_{b}^{\prime }$ is the inverse of the
scattering length. Since $k_{b}^{\prime }$ can be positive or negative, the
system may or may not possess a bound state.

As other examples of cases where the Case-Danilov prescription can be
applied, the reader can try it for the $s$-wave function of the hydrogen
atom~\cite{coutinho2008}, for the $s$-wave harmonic oscillator \cite%
{Correggi2008} in one, two and three dimensions or for the one-dimensional
hydrogen atom~\cite{coutinho08}.

\section*{Acknowledgements}
FABC acknowledges partial financial support from CNPq and Fapesp.
MA acknowledges partial financial support from CNPq.


\end{document}